\documentclass[twocolumn,aps,prd,preprintnumbers,showpacs,superscriptaddress,nofootinbib,amsmath,amssymb,floats,floatfix,showkeys,notitlepage,longbibliography]{revtex4-1}
\usepackage{orcidlink}
\usepackage{comment}
\usepackage{lipsum}
\usepackage{graphicx}
\usepackage{subfigure}
\usepackage{palatino}
\usepackage[utf8]{inputenc}
\usepackage[compat=1.1.0]{tikz-feynman}
\usepackage{subcaption}
\usepackage{amsmath,amssymb}
\usepackage{sans}
\usepackage{hyperref}
\hypersetup{colorlinks=true,linkcolor=blue,urlcolor=blue,citecolor=blue}
\usepackage[toc,page]{appendix}
\usepackage[normalem]{ulem}
\usepackage{adjustbox}
\usepackage{latexsym}
\usepackage{amsmath}
\usepackage{amssymb}
\usepackage{amsfonts}
\usepackage{dcolumn}
\usepackage{bm}
\usepackage{tikz}
\usepackage{bigints}
\usepackage{array,tabularx,multirow,booktabs}
\usepackage[tracking=true]{microtype}
\usepackage{multirow}
\SetTracking{}{500}
\SetTracking{encoding={*}, shape=sc}{40}
\UseRawInputEncoding 
\allowdisplaybreaks

 \newcommand{\bq}{\begin{equation}}
 \newcommand{\eq}{\end{equation}}
 \newcommand{\bqn}{\begin{equation}}
 \newcommand{\eqn}{\end{equation}}

\newcommand{\be}{\nopagebreak[3]\begin{equation}}
\newcommand{\ee}{\end{equation}}

\newcommand{\ba}{\nopagebreak[3]\begin{equation}}
\newcommand{\ea}{\end{equation}}

\NewDocumentCommand{\evalat}{sO{\big}mm}{%
  \IfBooleanTF{#1}
   {\mleft. #3 \mright|_{#4}}
   {#3#2|_{#4}}%
}

\begin{document} \sloppy
	\newcommand \nn{\nonumber}
	\newcommand \fc{\frac}
	\newcommand \lt{\left}
	\newcommand \rt{\right}
	\newcommand \pd{\partial}
	\newcommand \e{\text{e}}
	\newcommand \hmn{h_{\mu\nu}}
	
	\newcommand{\PC}[1]{\ensuremath{\left(#1\right)}} 
	\newcommand{\PX}[1]{\ensuremath{\left\lbrace#1\right\rbrace}} 
	\newcommand{\BR}[1]{\ensuremath{\left\langle#1\right\vert}} 
	\newcommand{\KT}[1]{\ensuremath{\left\vert#1\right\rangle}} 
	\newcommand{\MD}[1]{\ensuremath{\left\vert#1\right\vert}} 

\title{Black Hole Solutions in Dark Photon Models with Higher Order Corrections}

\author{Ali \"Ovg\"un \orcidlink{0000-0002-9889-342X}}
\email{ali.ovgun@emu.edu.tr}
\affiliation{Physics Department, Eastern Mediterranean University, Famagusta, 99628 North
Cyprus via Mersin 10, Turkiye.}

\author{Reggie C. Pantig \orcidlink{0000-0002-3101-8591}}
\email{rcpantig@mapua.edu.ph}
\affiliation{Physics Department, Map\'ua University, 658 Muralla St., Intramuros, Manila 1002, Philippines}

\begin{abstract}

In this work, we derive new analytic, static, symmetric black hole solutions in theories involving dark photons with minimal and higher-order magnetic dipole interactions. Starting from the effective non-relativistic potential between fermions mediated by a dark photon, we derive explicit corrections to the Schwarzschild geometry induced by dark photon and spin-dependent terms. These corrections alter the metric significantly at short distances, modifying the horizon radius, Hawking temperature, photon sphere, and consequently, the black hole shadow. Employing perturbative expansions, we provide analytic expressions for the deviations from the Schwarzschild solution, highlighting an exponential suppression controlled by the dark photon mass. Our results demonstrate that higher-order magnetic dipole interactions produce distinctive spin-dependent curvature terms, amplifying gravitational effects near the horizon. These findings provide a theoretical foundation for future phenomenological tests of dark photon models through gravitational wave astronomy and black hole imaging, while highlighting dark photons’ role as mediators of dark matter interactions that can influence structure formation and direct detection experiments.

\end{abstract}


\maketitle
\section{Introduction}

The nature of dark matter (DM) remains one of the central puzzles in modern particle physics and cosmology. 
Dark photons are hypothetical spin-1 gauge bosons associated with a hidden U(1) symmetry, proposed in many extensions of the Standard Model. They have attracted increasing interest as candidates for new physics, from dark matter to subtle quantum interference effects \cite{Holdom_1986}. On the conceptual side, Padavic-Callaghan has recently suggested that the classic double-slit experiment may be reinterpreted in terms of a dark-photon-mediated “phase memory” effect, wherein a hidden photon field carries and preserves the relative phase information of a quantum particle to produce the observed interference pattern \cite{Villas-Boas:2021rwm}. On the experimental side, searches for dark photon dark matter have yielded new results. An et al.~\cite{An:2025} report in situ measurements using NASA’s Parker Solar Probe to search for an ambient ultralight dark photon field in the solar corona. While no definitive signal was observed, their data set the strongest constraints to date on dark photons in the $\sim10^{-9}$ eV mass range (implying an extremely small kinetic mixing, $\chi \sim 10^{-13}$) and approach the sensitivity needed to detect the characteristic resonant conversion “bump” signature of dark photon dark matter. Such developments, bridging quantum laboratory phenomena and astrophysical observations, highlight the timely relevance of investigating dark photon models more deeply on theoretical grounds. Galactic dark matter halos can substantially influence both the generation and propagation of gravitational waves near black holes, underscoring the necessity of incorporating environmental dark matter effects in precise modeling of black hole signals \cite{Cardoso:2021wlq}. The presence of generic dark matter profiles around black holes alters their apparent image and modifies the characteristics of emitted gravitational waves, highlighting dark matter’s crucial role in interpreting black hole observations \cite{Figueiredo:2023gas}. Modeling the central spiky distribution of galactic dark matter as an Einstein cluster reveals how concentrated dark matter cores can affect black hole spacetime properties, emphasizing dark matter’s impact on black hole dynamics \cite{Maeda:2024tsg}. Treating dark matter as a fundamental vector field within the Einstein cluster framework provides an action principle that links dark matter distributions directly to black hole gravitational behavior, illustrating dark matter’s foundational importance in black hole theory \cite{Fernandes:2025lon}. Polar perturbations and energy flux analyses for black holes immersed in generic matter distributions demonstrate that dark matter’s surrounding structure significantly shapes the energy emission and stability of black hole spacetimes \cite{Speeney:2024mas}. Investigating a black hole embedded in the Dekel-Zhao dark matter profile illustrates how specific dark matter distributions can tailor the spacetime geometry and potentially observable signatures of black holes \cite{Ovgun:2025bol}.

While the gravitational effects of DM on galactic and cosmological scales are well established \cite{Planck2018}, its non-gravitational interactions are still largely unconstrained. A particularly well-motivated scenario posits a dark sector charged under a broken U(1) gauge symmetry, whose gauge boson $A'$ (the “dark photon”) kinetically mixes with the Standard Model hypercharge \cite{Holdom_1986,Pospelov2008}. If the dark photon mass $m_{A^\prime}$ lies in the MeV-GeV range, then dark matter particles may experience a new Yukawa-type force of finite range $r\sim 1/m_{A^\prime}$.

In the simplest realization, Dirac fermion DM carries a conserved dark U(1) charge $g_D$ and couples to the dark photon via a vector current.  At tree level, single $A'$ exchange induces the familiar spin-independent Yukawa potential  
\begin{equation}
V_{\rm min}(r)=-\frac{g_D^2}{4\pi}\,\frac{e^{-m_{A^\prime}r}}{r}\,,
\label{eq:Vmin}
\end{equation}
which reduces to a Coulomb potential in the massless limit $m_{A^\prime}\to0$ and has been studied extensively in the context of Sommerfeld-enhanced annihilation and dark bound‐state formation \cite{ArkaniHamed2009,Pospelov2008}.  Such a force can give rise to “dark atoms” \cite{Fayet2007} or “darkonium” \cite{MarchRussell2008}, with phenomenological implications for indirect detection and small-scale structure \cite{Buckley2010,TulinYu2018}.

If, instead, DM is a Majorana fermion or otherwise neutral under the dark U(1), renormalizable monopole couplings vanish but higher dimensional operators such as a magnetic dipole term may still arise.  A dipole coupling of the form   $(\mu_f/2\Lambda)\,\bar\chi\sigma^{\mu\nu}\chi\,F'_{\mu\nu}$ leads to a spin-dependent potential of tensor structure   \cite{Dobrescu:2006au,Fabbrichesi_2021}  
\begin{equation}
V_{\rm MD}(r)\approx -\frac{\mu_f^2}{\Lambda^2}\,\frac{e^{-m_{A^\prime}r}}{4\pi\,r^3}
\Bigl[\boldsymbol{\sigma}_1\!\cdot\!\boldsymbol{\sigma}_2-3(\boldsymbol{\sigma}_1\!\cdot\!\hat r)(\boldsymbol{\sigma}_2\!\cdot\!\hat r)\Bigr]\!,
\label{eq:Vmd}
\end{equation}
where \((\sigma_1,\sigma_2)\) are spin-dependent operators, analogous to the magnetic dipole-dipole interaction in electromagnetism \cite{Moxhay1983}.  This interaction is shorter‐range (scaling as $1/r^3$) and generically suppressed by $(\mu_f/\Lambda)^2\ll g_D^2$, but it can induce hyperfine splitting in dark bound states and give rise to spin-dependent scattering signatures in direct detection experiments \cite{Fan2013,DelNobile2014,Biswas:2022tcw}.

In light of these motivations, it is natural to ask how a dark photon field might manifest in strong-gravity environments such as black holes. Black holes in general relativity are famously described by only three classical parameters (mass, angular momentum, and electric charge), reflecting the stringent no-hair theorems. Introducing an additional U(1) gauge sector provides a potential new “hair” by endowing the black hole with a hidden charge or field configuration. For example, a massless dark photon would allow a black hole to carry a conserved hidden-sector charge, leading to an analog of the Reissner-Nordstrom solution that is invisible to ordinary electromagnetism. If the dark photon instead has a small mass, a rotating black hole may even support a surrounding cloud of dark photon excitations via superradiant instabilities \cite{Brito:2015}. These possibilities motivate extending black hole solutions to include a dark photon field. In this work, we take a step in this direction by constructing black hole solutions in an Einstein gravity theory augmented with a dark U(1) vector field (a “dark photon”). We examine how the presence of this hidden-sector gauge field modifies the black hole’s structure and dynamics, with the aim of elucidating any new phenomenological signatures or constraints that arise from the dark photon’s gravitational interactions.



Recent advances have underscored the possibility that dark photons not only serve as mediators within the dark sector but also induce detectable signatures through superradiance instabilities, Hawking radiation, or metric modifications near compact objects such as black holes. Superradiant amplification of ultralight bosonic fields in the vicinity of rotating black holes, a phenomenon deeply tied to the physics of dark photons, has catalyzed renewed efforts to connect quantum field theory in curved spacetime with dark sector phenomenology \cite{Cardoso_2018,Pierce:2018xmy,Fabbrichesi_2021,Caputo_2021}. Connection to the early universe string networks to the production of dark photon dark matter and constraints from black holes was also made \cite{Long_2019}.

The theoretical scaffold of dark photon physics is elegantly simple yet phenomenologically rich. In its most basic form, the interaction Lagrangian is described by kinetic mixing between the dark photon field $ A'_\mu $ and the hypercharge field of the Standard Model. This minimal scenario leads to Yukawa-like potentials between fermions mediated by dark photon exchange. However, the physics becomes notably more intriguing when non-minimal couplings, such as magnetic dipole, electric dipole, and anapole moments, are incorporated through dimension-5 and -6 operators. These terms introduce short-range corrections scaling as $ 1/r^3 $, and contribute spin-dependent forces with unique structural imprints \cite{Izaguirre_2016,Rogatko:2025zpc}.

Motivated by this framework, our work builds upon the recent non-relativistic treatment of fermionic scattering through minimal and magnetic dipole dark photon interactions. By computing the scattering amplitudes and transforming them into coordinate space via Fourier techniques, we derive the effective potential encompassing both Yukawa-type and spin-dependent corrections. Such corrections are not merely formal embellishments but exert real influence on gravitational field equations when the dark sector energy-momentum tensor is included as a source in Einstein's equations. Our analysis further explores how these additional interactions perturb the black hole metric, introducing modifications to the standard Schwarzschild-like geometry.

This effort aligns with several important developments in the literature. East and Huang \cite{East_2022} examined vortex dynamics in dark photon clouds formed by superradiance around spinning black holes, illuminating how nonlinear interactions shape their evolution. Siemonsen et al. \cite{Siemonsen_2023} expanded on the multimessenger consequences of dark photon electrodynamics, while Bhattacharyya et al. \cite{Bhattacharyya_2023} developed a worldline effective field theory incorporating dark photon effects into binary inspiral dynamics. Meanwhile, \cite{Cardoso_2018} demonstrated how black hole superradiance can be used as a sensitive probe for constraining the dark photon mass and coupling strengths.

From an astrophysical perspective, there is growing consensus that black holes immersed in dark matter halos may accrue effective charges or develop surrounding bosonic clouds through interactions mediated by dark photons \cite{Padilla_2024,Narzilloev_2021}. These configurations not only affect black hole spin evolution but may leave observable signatures in gravitational wave templates or electromagnetic counterparts. Our work builds a complementary line of inquiry by showing how higher-order corrections — particularly those from magnetic dipole interactions — deform the metric function $ f(r) $ and modify fundamental observables such as the event horizon radius and Hawking temperature.  Filho et al. (2024) derive and analyze the properties of charged black hole solutions modified by a Yukawa potential, revealing novel deviations from the Reissner–Nordstrom geometry \cite{Filho:2023abd,Filho:2024ilq}.  González et al. (2023) place observational constraints on Yukawa-type modifications to cosmology and demonstrate a compelling connection between these constraints and black hole shadow measurements \cite{Gonzalez:2023rsd}.

To this end, the present manuscript focuses on constructing a modified black hole solution within the dark photon framework, incorporating both minimal and magnetic dipole interactions. By solving the Einstein equations sourced by an effective energy density derived from the scattering potential, we derive analytic expressions for the metric components and their implications in the weak-field and near-horizon regimes. This approach offers a bridge between particle-level interactions in the dark sector and their macroscopic gravitational consequences.

Our contribution therefore lies in explicitly tracing how microscopic dark sector interactions mediated by a vector boson with magnetic dipole couplings translate into gravitationally measurable effects. It provides a consistent theoretical model applicable not only to phenomenological studies of dark matter detection but also to ongoing efforts to probe fundamental physics through black hole thermodynamics and gravitational wave observations.


\section{Dark Photon Corrections to Black Hole Geometry}
\label{secI}

We study how the spacetime geometry around a static, spherically symmetric black hole is modified by the presence of a dark photon field. We consider the general metric for such spacetime  given by
\begin{equation}
ds^2 = -f(r)\,dt^2 + \frac{dr^2}{f(r)} + r^2\left(d\theta^2+\sin^2\theta\, d\phi^2\right),
\end{equation}
where the metric function \(f(r)\) encodes gravitational corrections due to the dark photon sector.


The effective energy density due to dark photon contributions can be derived from the Laplacian of the potential corrections as \cite{Nicolini:2019irw,Filho:2024ilq}
\begin{equation}
\rho(r)=\frac{1}{4\pi}\Delta V(r),
\end{equation}
where the negative sign in potential terms indicates potential violations of classical energy conditions near the black hole.

Thus, the complete effective potential is 
\begin{equation}
V(r) = V_{\rm min}(r) + V_{\rm MD}(r)\,.
\end{equation}

where we consider minimal (Yukawa) and magnetic dipole (MD) terms, the potential as 

\begin{align}
V_{\text{min}}(r) &= -\frac{g_D^2}{4\pi}\frac{e^{-m_{A^\prime}r}}{r},\\[1mm]
V_{\text{MD}}(r) &\approx -\frac{\mu_f^2}{\Lambda^2}\frac{1}{4\pi}\frac{e^{-m_{A^\prime}r}}{r^3}\left[\boldsymbol{\sigma}_1\cdot\boldsymbol{\sigma}_2-3(\boldsymbol{\sigma}_1\cdot\hat{r})(\boldsymbol{\sigma}_2\cdot\hat{r})\right] \notag \\\notag &\equiv -\frac{\mu_f^2}{\Lambda^2}\frac{1}{4\pi}\frac{e^{-m_{A^\prime}r}}{r^3}S_{12}.
\end{align}

we find the corresponding Laplacians:
\begin{align}
\Delta V_{\text{min}}(r) &= -\frac{g_D^2 m_{A^\prime}^2}{4\pi}\frac{e^{-m_{A^\prime}r}}{r}+g_D^2\,\delta^{(3)}(\vec{r}), \\[2mm]
\Delta V_{\text{MD}}(r)&\approx -\frac{\mu_f^2}{\Lambda^2}\frac{1}{4\pi}\,e^{-m_{A^\prime}r}\left[\frac{m_{A^\prime}^2}{r^3}+\frac{4m_{A^\prime}}{r^4}+\frac{6}{r^5}\right]S_{12}.
\end{align}

Neglecting delta-function terms for \(r>0\), the energy density becomes
\begin{eqnarray}
\rho(r)=-\frac{g_D^2 m_{A^\prime}^2}{(4\pi)^2}\frac{e^{-m_{A^\prime}r}}{r}\notag \\-\frac{\mu_f^2 S_{12}}{\Lambda^2(4\pi)^2}\,e^{-m_{A^\prime}r}\left[\frac{m_{A^\prime}^2}{r^3}+\frac{4m_{A^\prime}}{r^4}+\frac{6}{r^5}\right].
\end{eqnarray}

Thus, the modified Einstein equation for the \(tt\)-component is
\begin{equation}
\frac{rf'(r)+f(r)-1}{r^2}=-8\pi\rho(r),
\end{equation}
which yields
\begin{eqnarray}
\frac{d}{dr}[r f(r)]=1+\frac{g_D^2 m_{A^\prime}^2}{2\pi}\,r e^{-m_{A^\prime}r}\notag \\+\frac{\mu_f^2 S_{12}}{2\pi\Lambda^2} e^{-m_{A^\prime}r}\left(\frac{m_{A^\prime}^2}{r}+\frac{4m_{A^\prime}}{r^2}+\frac{6}{r^3}\right).
\end{eqnarray}

Integrating and setting the integration constant as \(C=-2M\) to recover Schwarzschild behavior at large \(r\), we obtain
\begin{equation}
f(r)=1-\frac{2M}{r}-\frac{g_D^2 m_{A^\prime}}{2\pi}e^{-m_{A^\prime}r}-\frac{g_D^2}{2\pi}\frac{e^{-m_{A^\prime}r}}{r}+\frac{\mu_f^2 S_{12}}{2\pi\Lambda^2}\frac{I(r)}{r},
\end{equation}
where
\begin{equation}
I(r)=-2m_{A^\prime}^2 Ei(-m_{A^\prime}r)-\frac{m_{A^\prime} e^{-m_{A^\prime}r}}{r}-\frac{3e^{-m_{A^\prime}r}}{r^2},
\end{equation}
with \(Ei(z)\) being the exponential integral.


In the weak-field approximation, the metric function reduces to simpler analytical forms in two limits:

\textbf{1. Small-distance limit (\(m_{A^\prime}r\ll1\)):}\\[2mm]
At short distances, using expansions \(e^{-m_{A^\prime}r}\approx1\), and \(Ei(-m_{A^\prime}r)\approx\gamma+\ln(m_{A^\prime}r)\), one obtains
\begin{eqnarray}
f(r)\approx 1-\frac{2M}{r}+\frac{g_D^2(m_{A^\prime}r+1)}{2\pi r}\notag \\+\frac{\mu_f^2 S_{12}}{2\pi\Lambda^2}\left[\frac{2m_{A^\prime}^2(\gamma+\ln(m_{A^\prime}r))}{r}+\frac{m_{A^\prime}}{r^2}+\frac{3}{r^3}\right],
\end{eqnarray}
where \(\gamma\approx0.5772\) is Euler-Mascheroni constant.

\textbf{2. Large-distance limit (\(m_{A^\prime}r\gg1\)):}\\[2mm]
At large distances, one finds \(Ei(-m_{A^\prime}r)\approx -e^{-m_{A^\prime}r}/(m_{A^\prime}r)\), yielding
\begin{eqnarray}
f(r)\approx 1-\frac{2M}{r}+\frac{g_D^2}{2\pi}\frac{e^{-m_{A^\prime}r}(m_{A^\prime}r+1)}{r}\notag \\-\frac{\mu_f^2 S_{12}m_{A^\prime}e^{-m_{A^\prime}r}}{2\pi\Lambda^2 r^2}.\label{sol1}
\end{eqnarray}

\begin{figure}
    \centering
\includegraphics[width=0.48\textwidth]{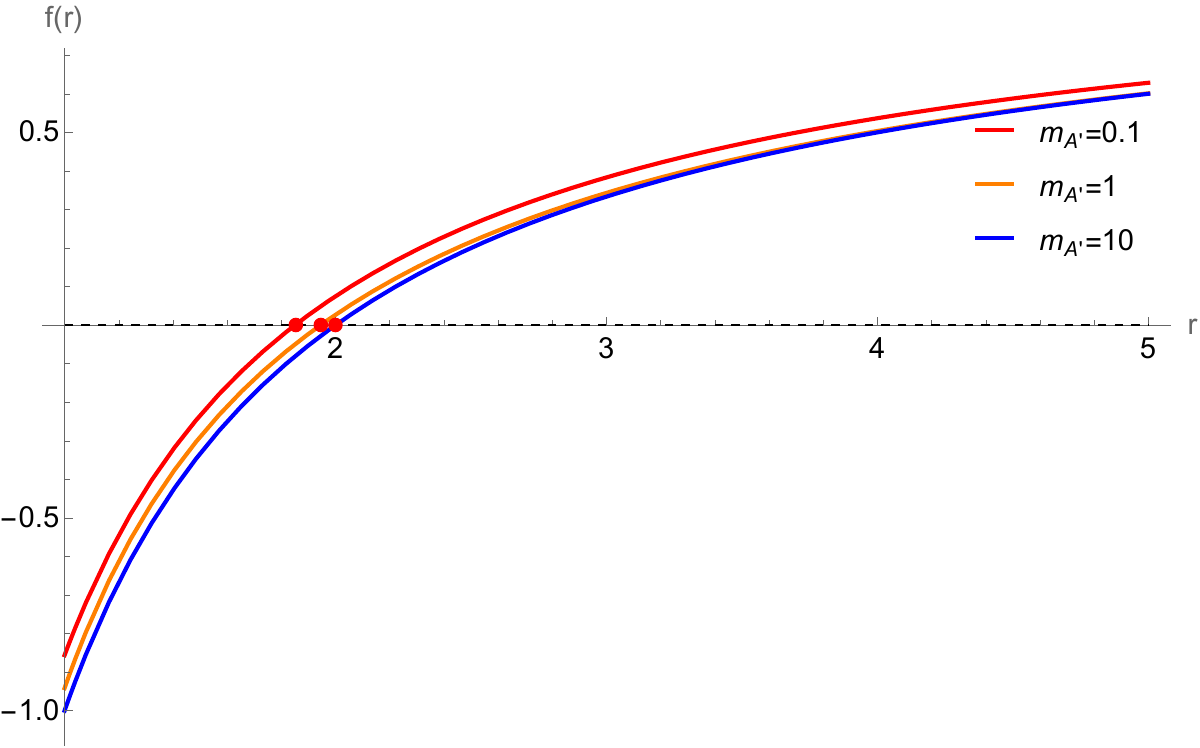}
    \caption{The plot shows the lapse function  $f(r)$  versus r for $M=g_D=\mu_f= S_{12}=1$.}
    \label{Fig:horizons}
\end{figure}


The event horizon \(r=r_+\) of the metric given in \ref{sol1} is defined by \(f(r_+)=0\) which is plotted in Fig. \ref{Fig:horizons}. At large distances, the horizon condition yields the black hole mass:
\begin{equation}
\begin{split}
M = \frac{r_+}{2}\Bigl[\,1 
  &+ \frac{g_D^2}{2\pi}\,\frac{e^{-m_{A^\prime}r_+}\,(m_{A^\prime}r_+ + 1)}{r_+} \\[6pt]
  &- \frac{\mu_f^2\,S_{12}\,m_{A^\prime}\,e^{-m_{A^\prime}r_+}}{2\pi\,\Lambda^2\,r_+^2}\Bigr]\,.
\end{split}
\end{equation}

The Hawking temperature, defined via surface gravity \(\kappa=f'(r_+)/2\), becomes
\begin{equation}
T_H=\frac{f'(r_+)}{4\pi},
\end{equation}
with
\begin{eqnarray}
f'(r_+)&=&\frac{2M}{r_+^2}-\frac{g_D^2}{2\pi}\left[m_{A^\prime}^2 e^{-m_{A^\prime}r_+}+\frac{e^{-m_{A^\prime}r_+}(m_{A^\prime}r_++1)}{r_+^2}\right] \notag\\ &+&\frac{\mu_f^2 S_{12}}{2\pi\Lambda^2}\left[\frac{m_{A^\prime}e^{-m_{A^\prime}r_+}}{r_+^2}+\frac{2e^{-m_{A^\prime}r_+}}{r_+^3}\right].
\end{eqnarray}

\begin{figure}
    \centering
\includegraphics[width=0.48\textwidth]{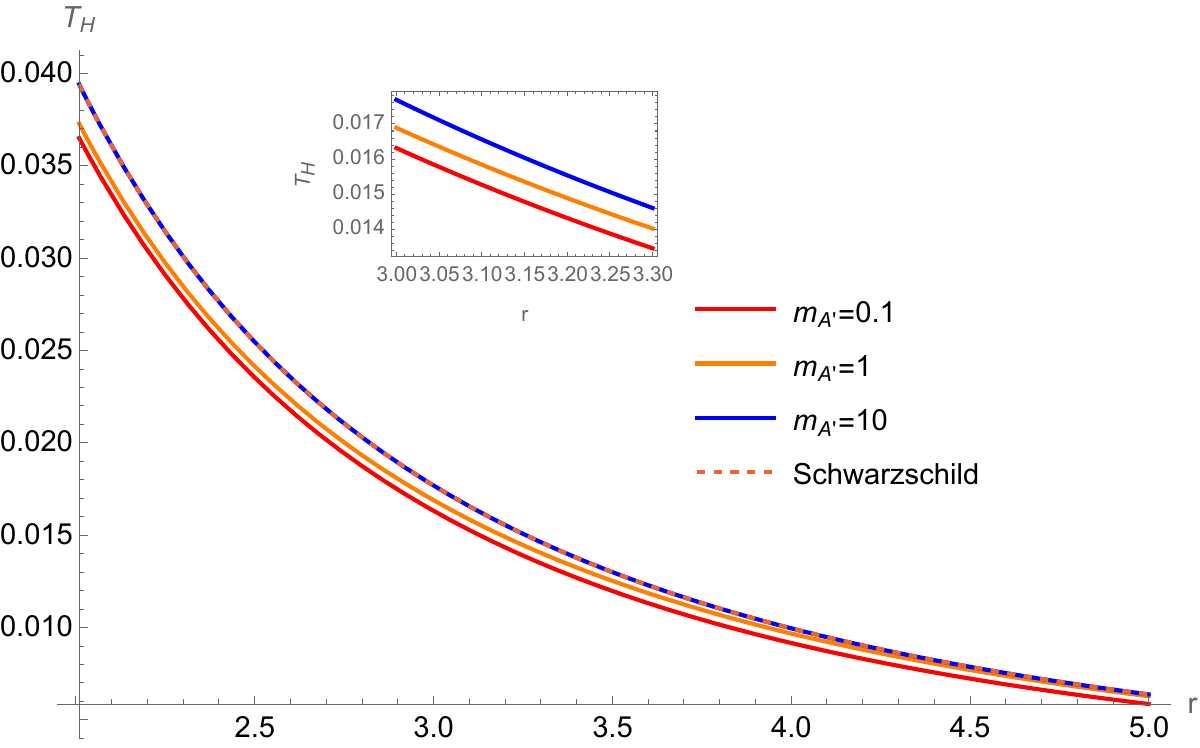}
    \caption{The plot shows the Hawking temperature  $T_H$  versus r for $M=g_D=\mu_f= S_{12}=1$.}
    \label{Fig:temperature}
\end{figure}

These expressions clearly demonstrate the physical effects of the dark photon corrections, indicating shifts in horizon radii and temperature. The Yukawa-like term modifies the Schwarzschild solution exponentially, while magnetic dipole terms introduce additional spin-dependent corrections that dominate at short distances shown in Fig. \ref{Fig:temperature}.

The Ricci scalar is given by
\begin{equation}
R = -\,f''(r) - \frac{4}{r}\,f'(r) + \frac{2\bigl[1 - f(r)\bigr]}{r^2}\,.
\end{equation}
Substituting Eqs.~\eqref{sol1}  leads to the Ricci scalar in the form
\begin{equation}
R = \alpha\,m^2\,\frac{e^{-mr}}{r}\,(3 - m r)
\;+\; C\,m^2\,\frac{e^{-mr}}{r^2}\,.
\label{eq:Ricci_final}
\end{equation}

with   
\begin{equation}
\alpha\equiv\frac{g_D^2}{2\pi},\quad
C\equiv\frac{\mu_f^2S_{12}\,m_{A^\prime}}{2\pi\Lambda^2},\quad
m\equiv m_{A^\prime}.
\end{equation}

The Kretschmann invariant reads
\begin{equation}
K = R_{\mu\nu\rho\sigma}R^{\mu\nu\rho\sigma}
= \bigl[f''(r)\bigr]^2
+4\Bigl(\tfrac{f'(r)}{r}\Bigr)^2
+4\Bigl(\tfrac{1 - f(r)}{r^2}\Bigr)^2\,,
\label{eq:K_general}
\end{equation}

 One may expand this to exhibit the Schwarzschild piece \(48M^2/r^6\) plus dark photon corrections.

The full contraction of the Ricci tensor is
\begin{eqnarray}
R_{\mu\nu}R^{\mu\nu}
=\frac{1}{2r^4}\Bigl[
f''^2\,r^4
+4\,f''\,f'\,r^3
+8\,f'^2\,r^2
\Bigr] \notag \\
+\frac{1}{2r^4}\Bigl[
8\,f\,f'\,r
+4\,f^2
-8\,r\,f'
-8\,f
+4
\Bigr].
\label{eq:RicciSq_part2}
\end{eqnarray}

To diagnose the central singularity, expand for \(r\ll m^{-1}\) using
\(\,e^{-mr}\approx1 - mr + \tfrac{1}{2}m^2r^2+\cdots\).  One finds
\begin{align}
f(r) &\sim -\frac{C}{r^2} + \frac{-2M+\alpha}{r} + \cdots,
\\
f'(r) &\sim +\frac{2C}{r^3} + \cdots,
\quad
f''(r)\sim -\frac{6C}{r^4} + \cdots.
\end{align}
Hence the leading divergences of the invariants are
\begin{align}
R(r) &\sim \frac{C\,m^2}{r^2} + \cdots \;\to\; +\infty,
\\
K(r) &\sim \frac{56\,C^2}{r^8} + \cdots \;\to\; +\infty,
\\
\mathrm{RicciSq}(r) &\sim \frac{20\,C^2}{r^4} + \cdots \;\to\; +\infty.
\end{align}
Thus \(r=0\) remains a true curvature singularity, with the strongest divergence in \(K\sim r^{-8}\).

\section{Estimating BHS in DPM with HOC Parameters} 
In static, spherically symmetric spacetimes, the apparent size of a black hole's shadow as seen by a distant observer is determined by the radius of the unstable circular null geodesic, commonly referred to as the \textit{photon sphere}. The condition for the existence of such a null circular orbit is given by the extremization of the effective potential, which yields the following equation for the photon sphere radius $r_{\rm ph}$ \cite{Claudel:2000yi}:
\begin{equation}
f'(r_{\rm ph})\,r_{\rm ph} \;-\; 2\,f(r_{\rm ph}) \;=\; 0 \,,
\label{eq:photon-sphere-condition}
\end{equation}
The corresponding shadow radius, or the critical impact parameter for photon capture, is then expressed as
\begin{equation}
R_{\rm sh} =\frac{r_{\rm ph}}{\sqrt{f(r_{\rm ph})}}\,.
\end{equation}

In the canonical Schwarzschild geometry, these yield the well-known results $r_{\rm ph}^{(0)} = 3M$ and $R_{\rm sh}^{(0)} = 3\sqrt{3}M$. However, deviations from Schwarzschild arise in a number of modified gravity frameworks or effective models coupling to exotic matter sectors, such as those mediated by Yukawa-type or exponential interactions. One such class of models introduces corrections to the lapse function via terms that decay exponentially with a scalar mass scale $m_{A^\prime}$, as explored in \cite{Odintsov:2024lid,McDermott:2017qcg}.

In the present analysis, we examine the theoretical shadow structure of a spherically symmetric black hole described by a deformation of the Schwarzschild metric wherein the gravitational potential is modified by exponential Yukawa-like corrections governed by parameters $\alpha$, $C$, and $m_{A^\prime}$. Writing
\begin{align}
f(r)&=1-\frac{2M}{r}
+\alpha\,\frac{e^{-m_{A^\prime}r}(m_{A^\prime}r+1)}{r}
-\;C\,\frac{e^{-m_{A^\prime}r}}{r^2}\!,
\nonumber \\
&\alpha\equiv\frac{g_D^2}{2\pi},\;
C\equiv\frac{\mu_f^2\,S_{12}\,m_{A^\prime}}{2\pi\Lambda^2},
\end{align}
we solve Eq.~\eqref{eq:photon-sphere-condition} perturbatively for $r = r_{\rm ph}$:
\begin{eqnarray}
r_{\rm ph}=3M+\delta,\qquad|\delta|\ll M\,,
\end{eqnarray}
to first order in \(\alpha\) and \(C\).  One finds
\begin{align}
r_{\rm ph}
&\approx 3M
-\frac{3\alpha M}{2}\,e^{-3m_{A^\prime}M}\bigl(3m_{A^\prime}^2M^2+3m_{A^\prime}M+1\bigr)\nonumber\\
&\quad +\,\frac{C}{6M}\,e^{-3m_{A^\prime}M}\,(3m_{A^\prime}M+4)\,.
\label{eq:rph-pert}
\end{align}
Substituting Eq.~\eqref{eq:rph-pert} into the definition of \(R_{\rm sh}\)
and expanding to the same order gives
\begin{align}
R_{\rm sh}=\frac{r_{\rm ph}}{\sqrt{f(r_{\rm ph})}}
&\approx 3\sqrt{3}\,M
\Bigl[1-\frac{\alpha \,e^{-3m_{A^\prime}M}(3m_{A^\prime}M+1)}{2M}\nonumber \\
&+\frac{C}{6M^2}\,e^{-3m_{A^\prime}M}\Bigr]\,.
\label{eq:Rsh-pert}
\end{align}
Therefore, the fractional deviations from the Schwarzschild values
\(\,r_{\rm ph}^{(0)}=3M,\;R_{\rm sh}^{(0)}=3\sqrt3\,M\)\, are
\begin{align}
\frac{\Delta r_{\rm ph}}{3M}
&\simeq -\frac{\alpha}{2}\,e^{-3m_{A^\prime}M}(3m_{A^\prime}^2M^2+3m_{A^\prime}M+1) \nonumber
\\ &+\frac{C}{6M^2}\,e^{-3m_{A^\prime}M}(3m_{A^\prime}M+4),
\end{align}
and
\begin{align}
    \frac{\Delta R_{\rm sh}}{3\sqrt3\,M}
&\simeq -\frac{\alpha}{2M}\,e^{-3m_{A^\prime}M}(3m_{A^\prime}M+1) \nonumber \\
&+\frac{C}{6M^2}\,e^{-3m_{A^\prime}M}\,.
\end{align}

\begin{figure}[t]
\includegraphics[width=\columnwidth]{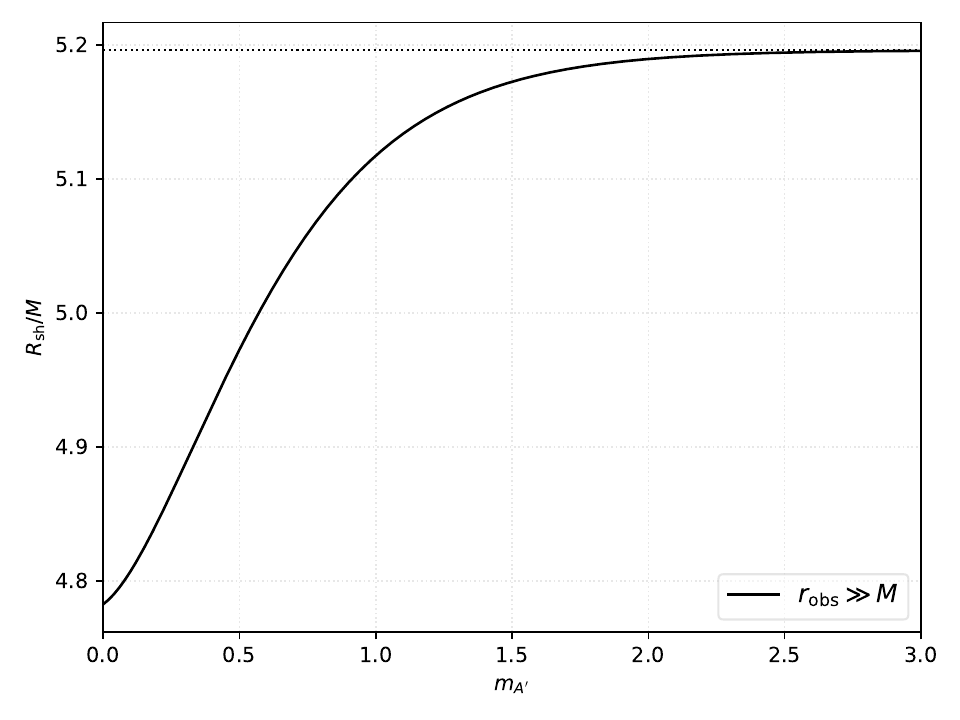}
\caption{Shadow radius \(R_{\rm sh}/M\) versus the dark photon mass \(mM\) for \(M=\alpha=1\) and $C=m$.  The horizontal dashed line is the Schwarzschild result \(3\sqrt{3}\).}
\label{fig:Rsh_vs_m}
\end{figure}

We plot the function $R_{\rm sh}$ in Fig. \ref{fig:Rsh_vs_m}. Notably, the shadow radius begins below the Schwarzschild limit and asymptotically approaches $3\sqrt{3}M$ from below as $m_{A^\prime} \to \infty$. This is a hallmark of theories where deviations are mediated by finite-range fields: the exponential factor $e^{-m_{A^\prime} r}$ ensures that for large mediator masses $m_{A^\prime}$, corrections decay rapidly and the geometry reduces to Schwarzschild at leading order. Conversely, for small $m_{A^\prime}$, deviations become non-negligible and the photon sphere shrinks, resulting in a smaller shadow size. Because of the overall exponential factor $e^{-3m_{A^\prime}M}$, deviations from the
Schwarzschild photon sphere and shadow radii are exponentially suppressed
for $m_{A^\prime}M\gtrsim1$, while for ultralight mediators ($m_{A^\prime}M\ll1$) they scale polynomially in $m_{A^\prime}$.  In either case the shifts are of order $\mathcal O(\alpha,C)$, and hence generically very small for weak
dark‐sector couplings.

Such behavior is consistent with analytic expectations derived in the perturbative regime. The correction to the photon-sphere radius and the shadow size takes the form \cite{Adler:2022qtb}:
\begin{equation}
    \frac{\Delta r_{\rm ph}}{3M},\, \frac{\Delta R_{\rm sh}}{3\sqrt{3}M} \sim \mathcal{O}(\alpha, C)\, e^{-3mM}\,P(mM)\,,
\end{equation}
where $P$ is a polynomial in $mM$. This exponential suppression implies that significant deviations can only be sustained for ultra-light mediators $m_{A^\prime} \ll M^{-1}$, or equivalently, long-range modifications to gravity.

Importantly, while the constants $\alpha$ and $C$ in our model encapsulate the strength of the exponential deformation, they are treated here as fixed but unspecified theoretical parameters. This avoids conflating the analysis with observational constraints from the Event Horizon Telescope or other phenomenological bounds, which would require both a fixed metric form and reliable knowledge of the source parameters (spin, inclination, emissivity, etc.). Our focus remains on the \textit{structure} of the theoretical deviation, rather than empirical data-fitting.

\label{shadow}

\section{Conclusions}
\label{conclusions}

We have derived new analytic solutions for static, spherically symmetric black holes in dark photon models-incorporating both minimal kinetic-mixing and higher-order magnetic dipole couplings and thoroughly analyzed their induced gravitational effects. By solving the Einstein field equations with the derived effective dark photon potential as a source, we identified explicit analytic corrections to the Schwarzschild metric. The resulting spacetime deviates significantly from the standard geometry, especially at short ranges, revealing distinct modifications in the horizon structure, Hawking temperature, and curvature invariants.

Our analysis explicitly quantifies how dark photon and spin-dependent magnetic dipole interactions introduce exponential and polynomial corrections to gravitational observables, respectively. These corrections lead to measurable changes in the photon sphere radius and shadow size, with characteristic exponential suppression governed by the dark photon mass. Remarkably, we found that magnetic dipole interactions induce stronger curvature singularities, with the Kretschmann scalar diverging faster than in the standard Schwarzschild case.

This study provides crucial theoretical insights into how microscopic dark sector couplings translate into macroscopic gravitational signatures. Such signatures are potentially observable through high-precision measurements in black hole imaging and gravitational wave detections. Thus, our findings bridge particle physics and gravitational phenomenology, showing the way for future observational tests of dark photon scenarios in extreme gravitational environments, and establishing a concrete connection between dark photons and dark matter phenomena by demonstrating how these hidden-sector gauge bosons can underpin observable signatures in cosmology and particle physics.

\acknowledgements
R. P. and A. \"O. would like to acknowledge networking support of the COST Action CA21106 - COSMIC WISPers in the Dark Universe: Theory, astrophysics and experiments (CosmicWISPers), the COST Action CA22113 - Fundamental challenges in theoretical physics (THEORY-CHALLENGES), the COST Action CA21136 - Addressing observational tensions in cosmology with systematics and fundamental physics (CosmoVerse), the COST Action CA23130 - Bridging high and low energies in search of quantum gravity (BridgeQG), and the COST Action CA23115 - Relativistic Quantum Information (RQI) funded by COST (European Cooperation in Science and Technology). A. \"O. also thanks to EMU, TUBITAK, ULAKBIM (Turkiye) and SCOAP3 (Switzerland) for their support.

\bibliography{ref.bib}

\end{document}